\documentclass[sigconf,anonymous=false,authorversion]{acmart}

\AtBeginDocument{%
  \providecommand\BibTeX{{%
    \normalfont B\kern-0.5em{\scshape i\kern-0.25em b}\kern-0.8em\TeX}}}

\setcopyright{acmlicensed}
\copyrightyear{2024}
\acmYear{2024}
\acmDOI{XXXXXXX.XXXXXXX}

\usepackage{contamination}
\begin{document}

\title[Training on the Test Model: Contamination in Ranking Distillation]{\bslink \ Training on the Test Model: \\
Contamination in Ranking Distillation}

\author{Vishakha Suresh Kalal}
\email{2885319k@student.gla.ac.uk}
\affiliation{%
  \institution{University of Glasgow}
  \city{Glasgow}
  \country{UK}
}
\author{Andrew Parry}
\authornote{Corresponding Author.}
\orcid{1234-5678-9012}
\email{a.parry.1@research.gla.ac.uk}
\affiliation{%
  \institution{University of Glasgow}
  \city{Glasgow}
  \country{UK}
}
\author{Sean MacAvaney}
\email{Sean.MacAvaney@glasgow.ac.uk}
\orcid{0000-0002-8914-2659}
\affiliation{%
  \institution{University of Glasgow}
  \city{Glasgow}
  \country{UK}
}

\renewcommand{\shortauthors}{Suresh Kalal et al.}

\begin{abstract}
  Neural approaches to ranking based on pre-trained language models are highly effective in ad-hoc search. However, the computational expense of these models can limit their application. As such, a process known as knowledge distillation is frequently applied to allow a smaller, efficient model to learn from an effective but expensive model. A key example of this is the distillation of expensive API-based commercial Large Language Models into smaller production-ready models. However, due to the opacity of training data and processes of most commercial models, one cannot ensure that a chosen test collection has not been observed previously, creating the potential for inadvertent data contamination. We, therefore, investigate the effect of a contaminated teacher model in a distillation setting. We evaluate several distillation techniques to assess the degree to which contamination occurs during distillation. By simulating a ``worst-case'' setting where the degree of contamination is known, we find that contamination occurs even when the test data represents a small fraction of the teacher's training samples. We, therefore, encourage caution when training using black-box teacher models where data provenance is ambiguous.
\end{abstract}


\maketitle

\section{Introduction}

Neural ranking models applying contextualized representations are frequently more effective than their statistical counterparts in ad-hoc ranking tasks~\citep[\textit{inter alia.}]{karpukhin:2020, formal:2022}. Learning from annotated examples enables a more precise approximation of relevance, though computation costs greatly increase. To partially reduce cost, a strong but large neural model can provide training data to a smaller student model in a semi-supervised fashion, often maintaining effectiveness while largely reducing latency~\cite{hofstatter:2020}. This process, known as knowledge distillation, allows the increasing size of models to be of minimal concern, as knowledge distillation allows small student models to capture the effectiveness of a strong teacher model when served in production. The use of Large Language Models (LLMs) as strong teachers exemplifies these benefits as high-quality semi-supervised data can be collected in a seemingly zero-shot fashion and distilled into highly effective smaller models~\citep{pradeep:2023, schlatt:2024b}. Many strong models are either fully closed-source~\citep{openai2024gpt4technicalreport} or solely open-weighted~\cite{touvron:2023, jiang:2023}; therefore, their training data is unknown. When researchers use these models as teachers in distillation, data provenance becomes important to precisely measure improvement due to novel contributions versus test set leakage. 

Because it is difficult to determine to what degree or in what form closed models have been exposed to a given test collection, we simulate a worst-case scenario to provide insights into test set leakage concerns for closed-source models.
This scenario directly optimizes teacher ranking models over training data contaminated with common test sets. We then employ these contaminated models as semi-supervised training signals in several common ranking distillation settings.

We observe significant improvements in the effectiveness of student models over contaminated teachers (compared to uncontaminated teachers), even when the test data leakage constitutes less than 0.1\% of the total training examples.
Therefore, we conclude that a semi-supervised training signal is sufficient to cause significant improvements over standard models on a contaminated benchmark. We find that both the explicit use of teacher output and RankNet-style order distillation lead to contamination. From this finding, we encourage caution when distilling from closed-source models and evaluating public benchmarks when data provenance is ambiguous. Even though we investigate through the lens of ranking tasks, our findings are closely related to many existing aspects of distillation in broader NLP due to the ever-increasing use of Plackett-Luce preference optimization in large language model alignment~\cite{wei:2022, chung:2022} and distillation~\cite{tunstall:2023}, which could suffer from similar contamination.

\section{Background}

Given a corpus $\mathcal{C} = \{d_i\}_{i=1}^{|\mathcal{C}|}$ and a query text $q$, a ranking model $f$ returns $k$ documents ranked by their relevance to $q$, \\ $\mathcal{R} = [d_i]_{i=1}^k, f(q, d_i) > f(q, d_{i+1})$. Neural ranking models learn to estimate relevance through a data-driven process. Training data is commonly composed of triples of text, $\mathcal{T_\text{train}} = \{Q, D^+, D^-\}$ where $D^+$ are human annotated relevant documents and $D^-$ are documents which are unlikely to be relevant and are often employed in a contrastive objective~\cite{nogueira:2019b, karpukhin:2020} to optimise parameters $\pi$. We focus on architectures over encoder-based models as such notions can be expanded to sequence-to-sequence models~\cite{nogueira:2020}. Generally, ranking models of the form $f(q, d;\pi)$ can be grouped as cross-encoders and bi-encoders. A cross-encoder encodes a joint representation of a query and document outputting a scalar value which can be interpreted as the probability $y$ of $d$ being relevant to $q$, formally $y = f(q, d;\pi), y \in \mathbb{R}$~\cite{nogueira:2019b}. This architecture affords deep interactions between query and document texts, and though effective, this joint representation must be computed online and is, therefore, inefficient. A bi-encoder instead uses the hidden states of an encoder model, encoding queries and documents before taking vector similarity as a surrogate for relevance estimation. The relevance of $d$ to $q$ can then be estimated as $y = \text{sim}(\pi(q), \pi(d)), y \in \mathbb{R}$ where $\text{sim}(\cdot)$ is any vector similarity function and $\pi(\cdot)$ encodes a text in latent space~\cite{nogueira:2019}. This process can be implemented efficiently by pre-computing document representations. 

Recall that in a contrastive setting, we use training data of the form $\mathcal{T_\text{train}} = \{Q, D^+, D^-\}$. In a distillation setting, we instead use explicit labels over data of the form $\mathcal{T_\text{train}} = \{Q, D\}$ with labels $Y = \{y_i : f(q_i, d_i;\pi), \forall q_i, d_i \in \mathcal{T_\text{train}}\}$. Therefore, our objective in a semi-supervised setting is to approximate the teacher's estimation of relevance. Originally, \citet{hinton:2015} proposed a point-wise mean-squared error criterion in a classification setting, finding that a smaller model could approximate the performance of a larger teacher; such a loss was first applied in a weakly supervised ranking setting by \citet{dehghani:2017}. However, employing task-specific inductive bias is highly effective in training ranking models; \citet{hofstatter:2020} proposed the marginMSE loss, which optimises the margin between positive and negative examples of a student and teacher. \citet{lin:2020} concurrently proposed the notion of tightly connected teachers using KL divergence to optimise score distributions over entire rankings. Several works now employ distillation methods purely using pairwise preferences~\cite{pradeep:2023, schlatt:2024b} in losses such as RankNet\citep{burges:2010} and variants of approximate nDCG~\cite{schlatt:2024b}

Though there is little work on the effect of contamination in ranking distillation as it is a recent concern, such contamination would most likely originate from closed-source models of which contamination has been more broadly investigated. Though several works look to approximate contamination~\citep{shi:2023, yang:2023} or infer training data~\citep{carlini:2020}, there is the inevitable caveat that we cannot know the true degree of contamination. Therefore, we consider a worst-case setting to be appropriate. The closest retrieval work to our own is that of \citet{frobe:2022}, which investigates contamination via similar queries in supervised settings; our work instead considers recent concerns in which true contamination can occur.

\section{Contamination via Distillation}

There are several real-world scenarios in which a teacher model could observe training data; the quantity and coverage of a given test collection may vary based on data provenance, which makes exact contributions of contamination to effectiveness ambiguous. To more concretely assess the effect of contamination in ranking distillation, we consider a worst-case scenario, being direct exposure to test data within a ranking optimisation objective, that is to say, for example, that we do not provide ID values and graded relevance but instead compose new training examples from test data. 

Formally, a test collection $\mathcal{T_\text{test}}$ over a corpus $\mathcal{C}$ will be composed of queries $Q$ and graded relevance judgements, which are a triple $\{q, d, g\}, g \in \mathbb{I}$ with $g$ denoting an annotated relevance grade for a query-document pair. For computing precision- and recall-based metrics, graded relevance judgements are binarized; we use this natural quantization to convert these query-document pairs into ``positive''  and ``negative'' examples as is used in ranking optimisation. For a relevance cutoff $r$, which is usually determined by the number of relevance grades, we produce a positive pool $D^+ = \{(q, d) : g >= r, \forall q, d, g \in \mathcal{T_\text{test}}\}$. Documents considered non-relevant in measure calculation are considered ``hard'' negatives and similarly collected. We take randomly sampled negatives from the corpus in cases of insufficient negative examples. As we model a worst-case scenario, we assume that the entire test collection is present in $\mathcal{T}_\text{train}$. 

\label{sec:negatives}
Our investigation controls for the addition of a small number of test examples in training data. For a given architecture and training regime, we produce two teachers, $\pi^*_\text{T}$ and $\pi_\text{T}$, with * denoting contamination of training data. A student architecture $\pi_S$ is trained under distillation objectives using the scores of each model, yielding $\pi^*_\text{S}$ and $\pi_\text{S}$.

\section{Evaluation}
We now outline the concrete steps in our investigation of the effects of contamination on ranking distillation. We look to provide evidence towards the following research questions:
\begin{itemize}
\item[\bf RQ-1] Does distillation of a contaminated teacher lead to inflated student effectiveness on a target collection? 
\item[\bf RQ-2] How does contamination under varying optimization settings affect effectiveness on a target collection? 
\item[\bf RQ-3] How does contamination affect effectiveness on other test collections?
\end{itemize}
We evaluate using several common benchmarks and neural architectures to best approximate common practice\footnote{Codebase: \href{https://github.com/Parry-Parry/ContaminatedDistillation/}{https://github.com/Parry-Parry/ContaminatedDistillation/}}.
\begin{table*}[!tb]
\centering
\scriptsize
\caption{Measuring the effect of training data contamination through distillation, rows denoted LCE are teacher models, with all other models being distilled from semi-supervised signals. \bslink \ denotes contamination of training data either by inclusion of test data or use of a contaminated teacher. Significance is denoted with respect to the standard teacher with \sig (paired t-test $p<0.05$).}
\label{tab:retrieal-effectiveness-ndcg}
\adjustbox{width=\textwidth}{
\begin{tabular}{ll@{\hspace*{.8em}}ccc@{\hspace*{.8em}}ccc@{\hspace*{.8em}}ccc}
\toprule
& & \multicolumn{3}{c}{Deep Learning 2019} & \multicolumn{3}{c}{Deep Learning 2020} & \multicolumn{3}{c}{TREC COVID} \\
\cmidrule(r@{.8em}){3-5}
\cmidrule(r@{.8em}){6-8}
\cmidrule(r@{.8em}){9-11}
 & Loss Criteria & nDCG@10 & MAP & R@100 & nDCG@10 & MAP & R@100 & nDCG@10 & MAP & R@100\\
\midrule
& LCE & 0.701\insig & 0.457\insig & 0.626\insig  & 0.700\insig & 0.486\insig & 0.719\insig & 0.590\insig & 0.161\insig & 0.147\insig \\
& LCE \bslink & 0.740\sig & 0.509\sig & 0.654\sig & 0.688\insig & 0.412\sig & 0.584\sig & 0.746\sig & 0.093\sig & 0.143\insig \\ 
\midrule
\parbox[t]{2mm}{\multirow{6}{*}{\rotatebox[origin=c]{90}{\textbf{Cross-En.}}}} & MarginMSE & 0.712\insig & 0.470\insig & 0.622\insig & 0.719\insig & 0.492\insig & 0.731\sig  & 0.610\insig & 0.160\insig & 0.146\insig \\ 
& MarginMSE \bslink & 0.728\sig & 0.495\sig & 0.639\sig & 0.719\insig & 0.494\insig & 0.729\insig & 0.690\sig & 0.088\sig & 0.143\insig \\
& KL Div. & 0.704\insig & 0.464\insig & 0.628\insig & 0.704\insig & 0.480\insig & 0.724\insig & 0.606\insig & 0.161\insig & 0.147\insig \\
& KL Div. \bslink & 0.707\insig & 0.471\sig & 0.628\insig & 0.713\insig & 0.491\insig & 0.723\insig & 0.657\sig & 0.084\sig & 0.143\insig \\
& RankNet & 0.698\insig & 0.458\insig & 0.605\sig & 0.705\insig & 0.484\insig & 0.723\insig & 0.604\insig & 0.163\insig & 0.150\insig \\
& RankNet \bslink & 0.715\insig & 0.492\sig & 0.632\sig  & 0.686\insig & 0.471\insig & 0.712\insig & 0.718\sig & 0.091\sig & 0.143\insig \\
\midrule
\parbox[t]{2mm}{\multirow{3}{*}{\rotatebox[origin=c]{90}{\textbf{Bi-En.}}}} & MarginMSE \bslink & 0.682\insig & 0.424\sig & 0.554\sig  &  0.637\sig & 0.425\sig & 0.685\sig  & 0.674\sig & 0.166\insig & 0.144\insig \\
& KL Div. \bslink & 0.665\insig & 0.405\sig & 0.549\sig & 0.657\sig & 0.442\sig & 0.691\sig  & 0.672\sig & 0.081\sig & 0.143\insig \\
& RankNet \bslink &0.670\insig & 0.433\insig & 0.582\sig & 0.633\sig & 0.408\sig & 0.692\sig  & 0.685\sig & 0.172\insig & 0.154\insig \\
\bottomrule
\end{tabular}
}

\end{table*}

\noindent\textbf{Datasets}
We primarily employ the MSMARCO passage collection composed of over 8.8 million segmented texts mined from Bing query logs~\citep{nguyen:2016}. We use this corpus for in-distribution (ID) evaluation employing the TREC Deep Learning (DL) 2019~\citep{craswell:2019} and 2020~\citep{craswell:2020} test collections consisting of 43 and 52 densely annotated queries, respectively. For out-of-distribution (OOD) contamination, we contaminate MSMARCO training data with the relevance judgements of the TREC COVID collection~\citep{roberts:2021} composed of medical queries and articles related to the COVID-19 pandemic.

\noindent\textbf{Models}
As a teacher architecture, we employ an ELECTRA-based cross-encoder~\citep{clark:2020}. Empirically, ELECTRA is more effective in several training settings than BERT in a cross-encoder setting~\citep{pradeep:2022}. As students, we employ both ELECTRA-based cross-encoders to model the effect of contamination on re-rankers and BERT-based bi-encoders~\citep{devlin:2019} to model a classic distillation setting~\citep{hofstatter:2020}. In all cases we re-rank BM25~\citep{robertson:1995} ($k_1=1.2$, $b=0.75$).

\noindent\textbf{Measures}
We primarily assess densely annotated test collections and, as such, evaluate precision-focused metrics nDCG@10 and MAP. Additionally, we measure Recall@100. For graded relevance ranges of 0 to 3, we use a relevance cutoff of 2, as is common practice in MSMARCO passage evaluation~\citep{craswell:2019}.

\noindent\textbf{Loss Criteria}
In training teacher models, we use cross-entropy loss in the form of Localized Contrastive Estimation (LCE)~\citep{gao:2021} exploiting ``hard'' negatives, which we take from BM25 and human-judged non-relevant texts (recall Section \ref{sec:negatives}). In a semi-supervised setting, we employ three loss criteria. We apply marginMSE~\citep{hofstatter:2020} and KL divergence minimisation~\citep{lin:2020}. Both of these criteria explicitly consider scalar deltas derived from a teacher's approximation of relevance; we additionally apply the RankNet loss~\citep{burges:2010}, which minimizes the error between a student and teacher \textit{ranking} instead of scores. This represents a common distillation method when using closed-source models~\citep{pradeep:2023, schlatt:2024b}, as logits are often unavailable in a list-wise sequence-to-sequence model~\cite{sun:2023}. To better approximate larger studies, we use multiple negatives when applying RankNet. Formally, for a training set of queries $Q$ we draw a negative sample from the top-$k$ ranked elements by $\pi_T$, $N = \{n | n \sim \mathcal{R}_\pi(q), \forall q \in Q\}$ following prior works~\cite{pradeep:2023, schlatt:2024b}.

\noindent\textbf{Training Regime}
We train each model for a single epoch in the case of a group size of 2 (1 negative, approximately $6e^{6}$ gradient updates) and $3e6$ steps for RankNet distillation due to computational constraints when using a group size of 8. We employ a global batch size of 32 queries in all settings. We use the AdamW optimizer with a learning rate of $1e^{-5}$ and a 10\% linear warm-up and decay schedule. The training regime bar contamination is identical in all settings apart from the loss criteria applied as described above. We train each model in mixed float-16 precision on an RTX 4090 GPU.
\begin{table}[tb]
    \centering
    \caption{Comparing ranking effectiveness of student models in terms of nDCG@10 with different contamination sources. Each model was trained using KL-Divergence minimisation with its respective teacher.}
    \label{tab:compare}
    \begin{tabular}{l@{\hspace*{.8em}}ccc}
        \toprule
        \textbf{Contamination Source} & \textbf{DL2019} & \textbf{DL2020} & \textbf{COVID} \\
        \midrule
        None & 0.704 & 0.704 & 0.606 \\
        Deep Learning 2019 & 0.707 & \textbf{0.725} & 0.612 \\
        Deep Learning 2020 & \textbf{0.717} & 0.713 & 0.623 \\
        COVID & \textbf{0.717} & 0.712 & \textbf{0.657}\\
        \bottomrule
    \end{tabular}
   
\end{table}

\section{Results and Discussion}

\noindent\textbf{Distillation Yields Downstream Contamination.}
The core question investigated in this work is the degree to which contamination can transfer in downstream training processes. Similarly to \citet{frobe:2022}, we observe inflation of effectiveness when directly contaminating training data. Observe that the contaminated teacher (LCE \bslink) improves significantly in nDCG@10 and MAP in two of three datasets; however, we find that in the case of DL2020, contamination can degrade performance in a standard training setting\footnote{We leave the investigation of this case to future work as we find that this teacher signal is still sufficient to improve in downstream distillation.}. Addressing \textbf{RQ-1}, observe that a contaminated distillation can frequently reach parity with or exceed the effectiveness of a standard training process and standard distillation. Particularly in the case of contamination with OOD examples, a cross-architecture distillation can exceed the effectiveness of a standard cross-encoder by over 10 points of nDCG@10 without observing test data. However, we see that the reduced MAP scores of a contaminated OOD teacher are transferred to students in some cases, which could suggest over-fitting.
Furthermore, we generally observe an improvement in recall greater than that of a standard distillation setting. We also observe that in an OOD setting, a contaminated bi-encoder can significantly improve over a standard model in terms of nDCG@10. Addressing \textbf{RQ-2}, when distillation is applied to homogeneous students and teachers, the largest disparities in distilled effectiveness are observed over RankNet distillation. In this case, text ordering is used instead of explicit scores. This setting is commonly used to distil closed-source models but may lead to greater contamination; this finding is concerning given the increasing prevalence of this approach, and we would look to more deeply investigate whether such a signal is more effective in modelling a teacher's relevance distribution in future work.

\noindent\textbf{Cross-Contamination Can Yield Greater Effectiveness.}
Concerning \textbf{RQ-3}, in Table \ref{tab:compare}, the effectiveness of ID and OOD contamination is contrasted across each dataset in terms of nDCG@10. We see that OOD contamination can improve over in-distribution contamination even when evaluating the in-domain test set from which contaminated examples were taken, as surprisingly, the OOD contamination outperforms the explicit contamination similarly to the explicit training investigated by \citet{frobe:2022}. This finding is concerning given that even if we were to move from a collection as ubiquitous as TREC Deep Learning, in providing multiple hard examples from test corpora, we may still largely influence downstream effectiveness across other collections; in that sense, any contamination should be treated with caution not simply that of an in-distribution collection.

\section{Conclusions and Future Work}
As closed-source models are increasingly present in academic work, we investigate a worst-case scenario in which explicit test set contamination occurs during model training. In contaminating the training data of a model with test data before training a student model, we observe that in several semi-supervised settings, clear inflation of downstream effectiveness occurs, which, in some cases, exceeds the effectiveness of a standard teacher model. From these findings, we suggest caution in comparing systems where semi-supervised data provenance is ambiguous, as sources of effectiveness may not be from a novel approach but from any number of contamination sources.

\bibliographystyle{ACM-Reference-Format}
\bibliography{sample-base}

\end{document}